\documentstyle[prl,aps,epsf,multicol]{revtex}
\begin{document}
\draft

\def\i{\imath\,}
\def\ih{\frac{\imath}{2}\,}
\def\undertext#1{\vtop{\hbox{#1}\kern 1pt \hrule}}
\def\ra{\rightarrow}
\def\lfa{\leftarrow}
\def\ua{\uparrow}
\def\da{\downarrow}
\def\Ra{\Rightarrow}
\def\lra{\longrightarrow}
\def\ler{\leftrightarrow}
\def\lrb#1{\left(#1\right)}
\def\O#1{O\left(#1\right)}
\def\EV#1{\left\langle#1\right\rangle}
\def\tr{\hbox{tr}\,}
\def\trb#1{\tr\lrb{#1}}
\def\dd#1{\frac{d}{d#1}}
\def\dbyd#1#2{\frac{d#1}{d#2}}
\def\pp#1{\frac{\partial}{\partial#1}}
\def\pbyp#1#2{\frac{\partial#1}{\partial#2}} 
\def\pd#1{\partial_{#1}}
\def\br{\\ \nonumber & &}
\def\brr{\right. \\ \nonumber & &\left.}
\def\inv#1{\frac{1}{#1}}
\def\be{\begin{equation}}
\def\ee{\end{equation}}
\def\bea{\begin{eqnarray}}
\def\eea{\end{eqnarray}}
\def\ct#1{\cite{#1}}
\def\rf#1{(\ref{#1})}
\def\EXP#1{\exp\left(#1\right)} 
\def\INT#1#2{\int_{#1}^{#2}} 
\def\LHS{left-hand side }
\def\RHS{right-hand side }
\def\COM#1#2{\left\lbrack #1\,,\,#2\right\rbrack}
\def\AC#1#2{\left\lbrace #1\,,\,#2\right\rbrace}

\title{Detecting fractions of electrons in the high-$T_c$ cuprates}
\author{T. Senthil and Matthew P.A. Fisher}
\address{
Institute for Theoretical Physics, University of California,
Santa Barbara, CA 93106--4030
}

\date{\today}
\maketitle

\begin{abstract}
We propose several tests of the idea that the electron is fractionalized in the underdoped
and undoped cuprates. These include the ac Josephson effect, and tunneling into 
small superconducting grains in the Coulomb blockade regime. In both cases, we argue that 
the results are qualitatively modified from the conventional ones if the insulating tunnel
barrier is fractionalized. These experiments directly detect the possible existence of the 
chargon - a charge $e$ spinless boson - in the insulator. The effects 
described in this paper provide a means to probing whether the undoped cuprate 
(despite it's magnetism) is fractionalized. Thus, the experiments discussed here are complementary
to the flux-trapping experiment we proposed in our earlier work\cite{toexp}.

\end{abstract}
\vspace{0.15cm}


\begin{multicols}{2}
\narrowtext

\section{Introduction}
Superconductivity occurs in solids when a charged excitation with 
Bose statistics condenses. The electrons in a solid are fermions, 
and cannot directly condense to produce superconductivity. A 
well-known solution 
to this difficulty is to pair electrons together into 
Cooper pairs. The Cooper pairs being bosons can then 
condense giving rise to superconductivity. An alternative solution
is to splinter the electron\cite{PWA,KRS,LN,z2short} into two pieces, 
thereby liberating it's charge from it's Fermi statistics. The 
resulting charged boson can then condense leading to 
superconductivity\cite{KRS,z2long,z2short}. 
Remarkably, the superconductor so obtained is in the same phase\cite{SK,SNL,z2long} as a 
BCS superconductor obtained by condensing Cooper pairs. In other words, 
both routes to superconductivity lead to the same final destination. 
In conventional metals, the occurence of superconductivity is 
attributed to the presence of Cooper pairing of the Landau 
quasiparticles of the normal Fermi liquid 
state due to attractive interactions 
arising from phonons. In the cuprate high-$T_c$ superconductors,
on the other hand, it may well be that the superconductivity occurs via 
the splintering of the electron. Some evidence for this is 
provided by angle-resolved photoemission experiments\cite{arpesrev} which do not
see any evidence for Landau quasiparticles in the normal state.

The quantization of electromagnetic flux in units of 
$hc/2e$ is usually taken as evidence of the presence of Cooper pairing
in superconductors.  However, in the ``non-pairing" fractionalization route to superconductivity - which is driven by the condensation
of a charge $e$ chargon - $hc/2e$ flux quantization is nevertheless
possible due to 
the presence of topological 
``vortex-like'' excitations\cite{ReCh,SK,z2long} in the normal state - dubbed the visons. 
Indeed, the existence of visons as gapped excitations is crucial\cite{z2long} for the 
electron to be able to fractionalize at all. Recently, we proposed an 
experiment\cite{toexp,topth} that could directly detect the visons in the normal state
thereby providing a direct test of the idea that the electron fractionalizes 
in the cuprates.

In this paper, we propose several other tests of the idea 
that the electron is fractionalized in the non-superconducting state. 
These explore parts of the 
cuprate phase diagram that are different from those explored by the 
vison detection experiment. 

We first examine the Josephson effect in 
superconductor-insulator-superconductor junctions. In the classic
ac Josephson effect,    
a dc voltage $V$ applied to this junction leads to oscillations 
of the current at a frequency $\omega = \frac{2eV}{\hbar}$. This 
fundamental result has been used to set the standard measure of 
the unit of voltage\cite{volt}. The factor of $2$ indicates that the tunneling current 
is carried by charge $2e$ Cooper pairs. In contrast to this classic effect, 
we argue that if the insulator in the junction is fractionalized, there  
will in addition be oscillations at a frequency $\omega = \frac{eV}{\hbar}$. 
The ratio of the amplitudes of the oscillations at the two different
frequencies depends on the charge gap and the vison gap in the
fractionalized insulator. A good candidate to maximize the amplitude of the 
$\frac{eV}{\hbar}$ oscillation is the {\em undoped} cuprate insulator.
The undoped cuprates are antiferromagnetic Mott insulators. However, as
pointed out in Ref. \cite{NLII,topth}, the fractionalization of the electron 
could coexist with the magnetism. Observation of $\frac{eV}{\hbar}$ 
oscillations will
establish experimentally the fractionalization of the electron in the 
undoped cuprate. 

In recent years, a number of experiments\cite{cb} probing tunneling into  
small grains of conventional low-$T_c$ materials 
have shown an ``even-odd'' effect: the tunneling conductance
has a periodic sequence of peaks  as a function of the total charge on the grain. 
The period is {\em twice} the electron charge - this can be interpreted 
as due to Cooper pairing of electrons in the superconductor. We argue that this result
will be modified if the insulating barrier in the tunnel junction is 
fractionalized. Specifically, we consider the situation where the tunneling occurs from 
superconducting leads through an insulating tunnel barrier to a small 
superconducting grain. If the insulator is fractionalized, it becomes possible 
for chargons to tunnel through. Consequently, the tunneling conductance would have a 
periodic sequence of peaks with period set by $e$ rather than $2e$. 

Another test of the fractionalization scenario for the underdoped cuprates
was pointed out a long time ago by Sachdev, Nagaosa, and Lee\cite{SNL}. They observed that
a superconductor that descends from a fractionalized insulator has regimes in which
the energy cost of an $hc/e$ vortex is smaller than two isolated $hc/2e$ vortices. 
Thus observation of stable $hc/e$ vortices in the superconducting phase
would be an indirect test of the fractionalization in the ``normal" state. 
Here, following their ideas, and using currently available data, we provide a 
rough estimate of the region of stability of the $hc/e$ vortex.   

\section{$Z_2$ gauge theory}
We begin by very briefly reviewing the theory of the fractionalized
insulator. The excitations in 
the fractionalized insulator are a charge $e$ spinless boson (the chargon)
and a chargeless spin-$1/2$ fermion (the spinon). In addition, 
there is a gapped $Z_2$ vortex excitation (the vison). The 
details of the spin physics in this fractionalized insulator are 
not important for our purposes: in particular, the insulator could 
have magnetic long range order. In the context of the cuprates, 
this is significant. The undoped insulator certainly has Neel magnetic order,
but may nevertheless also be fractionalized\cite{NLII,topth}. 

A very convenient theoretical language to described the fractionalized insulator
is provided by the $Z_2$ gauge theory formulation developed in Refs. \cite{z2long}.
The $Z_2$ formulation recasts a general class of interacting electron models
as a theory of chargons and spinons minimally coupled to a fluctuating 
$Z_2$ gauge field. 
A Hamiltonian version of the $Z_2$ gauge theory is:
\bea
H & = & H_c + H_{\sigma} + H_s, \\
H_c & = & -\sum_{<rr'>} t_{rr'} \sigma^z_{rr'} \left(b^{\dagger}_r b_{r'}
+ h.c \right) + U \sum_r \left( N_r - 1 \right)^2, \\
H_{\sigma} & = & -K\sum_{\Box} \prod_{\Box} \sigma^z_{rr'} - h \sum_{<rr'>} \sigma^x_{rr'}, \\
H_s & = & -\sum_{<rr'>} \sigma^z_{rr'}\left[t_s \left(f^{\dagger}_r f_{r'} + 
h.c \right) \right. \nonumber \\
+ & & \Delta_{rr'} \left. \left(f_{r\ua} f_{r'\da} - f_{r \da} f_{r' \ua}  + h.c \right) \right]
+ H_{int}[f].
\eea   
Here $b^{\dagger}_r$ creates a chargon at site $r$ while $f^{\dagger}_{r\alpha}$
creates a spinon with spin $\alpha = \ua, \da$ at site $r$. The 
operator $N_r = b^{\dagger}_rb_r$ measures the number of bosons at site $r$. For simplicity, we have 
specialized to half-filling, {\em i.e}, to an average of one boson per site.
The constant $\Delta_{rr'}$
contains the information about the pairing symmetry of the spinons. For the present, 
we assume that $\Delta_{rr'}$ has $d_{x^2 - y^2}$ symmetry. 
The term $H_{int}$ in the spinon
Hamiltonian represent four spinon interaction terms which 
could induce antiferromagnetic
ordering of the spin.    
The $\sigma^z_{rr'},
\sigma^x_{rr'}$ are Pauli spin matrices that are defined on the links of the lattice. 
The $\sigma^z_{rr'}$ may be thought of as $Z_2$ gauge fields. The full
Hamiltonian is invariant under the $Z_2$ gauge transformation $b_r \ra -b_r, 
f_r \ra -f_r$ at any site $r$ of the lattice accompanied by letting $\sigma^z_{rr'}
 \ra -\sigma^z_{rr'}$ on all the links connected to that site. 
This Hamiltonian must be supplemented with  
the constraint equation
\be
G_r = \Pi_{r' \in r}\sigma^x_{rr'} e^{i\pi \left(f^\dagger_r f_r + N_r\right)} = 1.
\ee
Here the product over $\sigma^x_{rr'}$ is over all links that emanate
from site $r$.
The operator $G_r$, which commutes with the
full Hamiltonian, is the generator of the local $Z_2$ gauge symmetry. Thus the 
constraint $G_r =1$
simply expresses the condition that the physical states in the Hilbert space are those that are gauge
invariant.   

The fractionalized insulating phase is described as the deconfined phase
of this gauge theory. This is obtained when $K >> h, U >> t_{rr'}$. 
On the other hand, the conventional 
superconductor is described as a phase in which the chargons have condensed.
This is obtained when $t_{rr'} >> U$, or alternately by doping away from half-filling. 
Note that the ``pairing''  symmetry of the 
superconductor is determined by $\Delta_{rr'}$.   

\section{Josephson effect}
\begin{figure}
\epsfxsize=3.3in
\centerline{\epsffile{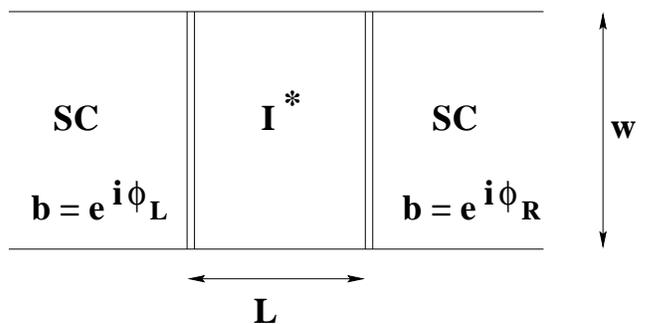}}
\vspace{0.15in}
\caption{Schematic of the superconductor-insulator-superconductor junction.
Here $I^*$ refers to the fractionalized insulator. $\phi_L$ and $\phi_R$
are the phases of the {\em chargons} in the left and right superconductors.}
\vspace{0.05in}
\label{sis}
\end{figure}

Now consider a superconductor-insulator-superconductor junction such as that shown in 
Fig \ref{sis}. We assume that the insulator is fractionalized. As the spin physics is 
irrelevant for the following discussion, we drop the spinon dependant term
in the Hamiltonian above, and just focus on $H_c + H_{\sigma}$. We assume 
that $t_{rr'} = t >> U$ in both superconducting regions, and $t_{rr'} = t' << U$ 
in the insulating region.
We also assume that $K >> h$ in the insulating region. 

Inside the superconducting regions, we may safely ignore vortices in the phase of the 
chargon field. In particular, we may set $\sigma^z_{rr'} = +1$ for all links inside 
the superconducting regions. The phase of the chargon field is fixed inside 
both superconducting regions. We write $b_r \approx e^{i\phi_r}$ with $\phi_r = \phi_L$
in the superconductor to the left, and $\phi_r = \phi_R$ in the superconductor 
to the right. We wish to derive the coupling between $\phi_L$ and $\phi_R$ due to the 
insulating region between the two superconductors. As $t' << U$ in the fractionalized
insulator, we may perturbatively integrate out the chargon degrees of freedom. 
For an insulating region of size $L$ (measured in units of the 
lattice spacing), the lowest order term coupling $\phi_L$
and $\phi_R$ is obtained in the $L$'th order of perturbation theory. The 
resulting effective Hamiltonian is 
\be
H_{eff} = -t_{eff}\left(\sum_C \prod_C \sigma^z_{rr'} \right) \cos(\phi_L - \phi_R) 
+ H_{\sigma},
\ee
with $t_{eff} \sim t' \left(\frac{t'}{U}\right)^{L-1}$. Here $C$ denotes a straight 
line path from any point on the left interface to the corresponding point 
on the right interface.
To obtain the Josephson coupling, we further need to integrate out the 
$Z_2$ gauge degrees of freedom. Consider the first term of the effective
Hamiltonian $H_{eff}$ above as a perturbation to $H_{\sigma}$. 
To leading order, we may replace $\prod_C \sigma^z_{rr'}$ 
by it's average evaluated with $H_{\sigma}$. This average
is readily found for small $h/K$ (which is the appropriate limit in the fractionalized
insulator). 
In the limit that $h = 0$, we may set $\sigma^z_{rr'} = 1$. For small $h/K$, 
each $\sigma^z_{rr'}$ has an amplitude $h$ to be negative while the energy cost 
for this fluctuation is order $K$. Thus the average value of the product of
$\sigma^z_{rr'}$ over any path $C$ will decay exponentially with the
length of the path:
$< \prod_C \sigma^z_{rr'} > \sim e^{-\lambda_v L}$ with $\lambda_v \sim h/K$.
Thus, to leading order, we get the coupling
\be
E^{(1)}_J = -t_1 \cos(\phi_L - \phi_R).
\ee
Here $t_1 \sim U w e^{-(\lambda_c + \lambda_v)L}$ with $\lambda_c = ln(U/t')$,
and $w$ is the lateral width of the junction. (Strictly speaking, 
this result assumes the thermodynamic limit in the lateral direction, {\em i.e}, large
$w$ - see below.) 

Physically, this represents a direct coupling between the phases of the chargons
in the two superconductors due to coherent tunneling of {\em chargons} through the 
intervening insulator. It is potentially also important to include the 
effect of coherent tunneling of {\em Cooper pairs} between the two superconductors. 
This is obtained at order $2L$ in the perturbation theory when integrating
out the chargon fields. The result
is
\be
E^{(2)}_J = -t_2 \cos(2\phi_L - 2\phi_R),
\ee
with $t_2 \sim wt'\left(\frac{t'}{U}\right)^{2L -1} = wU e^{-2\lambda_c L}$.
Thus the full Josephson coupling between the two superconductors is
\be
\label{Jos}
E_J =  -t_1 \cos(\phi_L - \phi_R) -t_2 \cos(2\phi_L - 2\phi_R).
\ee
We emphasize that $\phi_L, \phi_R$ represent the chargon phase in the 
two superconductors. The Cooper pair phase is twice the chargon phase. 
The second term is therefore the ``standard'' Josephson 
coupling while the first is novel, and arises due to the possibility of
coherent tunneling of chargons through the fractionalized insulator. 
The ratio between the amplitudes of the chargon and Cooper pair 
tunneling terms is
\be
\frac{t_1}{t_2} \sim e^{(\lambda_c - \lambda_v)L}.
\ee
Deep inside the fractionalized insulating phase, we 
have $\lambda_c \sim ln\left(\frac{U}{t'}\right) >1$, and 
$\lambda_v \sim \frac{h}{K} << 1$. Thus $t_1$ will then dominate over
$t_2$. In general, the optimal situation to maximize the ratio
$t_1/t_2$ is to have an insulating barrier with a large charge gap ({\em i.e}
large $U/t'$)
and a large vison gap ({\em i.e} large $K/h$). 

In the context of the cuprates, this suggests that the best prospects
for observing coherent chargon tunneling will occur if the 
insulating barrier is made of the {\em undoped} cuprate. This 
is as deeply insulating as is possible in the cuprates. Further, 
the vison gap (estimated to be of order the pseudogap temperature)
is perhaps the largest in the undoped material. 

From now on, we assume that $t_1 >> t_2$. The form of the 
Josephson coupling with only the chargon tunneling term has the 
immediate consequence that the ac Josephson frequency
will be $\frac{eV}{\hbar}$ which differs by a factor of two
from the conventional one. This is a direct probe of the charge
of the boson that tunnels coherently between the two 
superconductors. Observation of such $\frac{eV}{\hbar}$ oscillations in the
ac Josephson effect will
prove that the undoped cuprate is fractionalized (despite it's Neel 
long range order). 

In passing, we note that even in the conventional ac Josephson effect
as probed by the Shapiro steps in an irradiated junction, for instance, 
subharmonic oscillations at frequency $\frac{2eV}{n\hbar}$ with integer $n$ are 
present\cite{tinkh}. 
These occur due to the possibility of absorption of $n$ photons
by the tunneling Cooper pairs. However, for small intensity of the 
radiation, the amplitude for the processes with $n > 1$ is substantially
smaller than for $n = 1$. We may then safely ignore the possibility of 
multi-photon absorption. As the intensity of the radiation is increased
on the fractionalized junction, 
multi-photon absorption should also become possible leading to subharmonic oscillations
at frequencies $\frac{eV}{n\hbar}$.

\begin{figure}
\epsfxsize=3.3in
\centerline{\epsffile{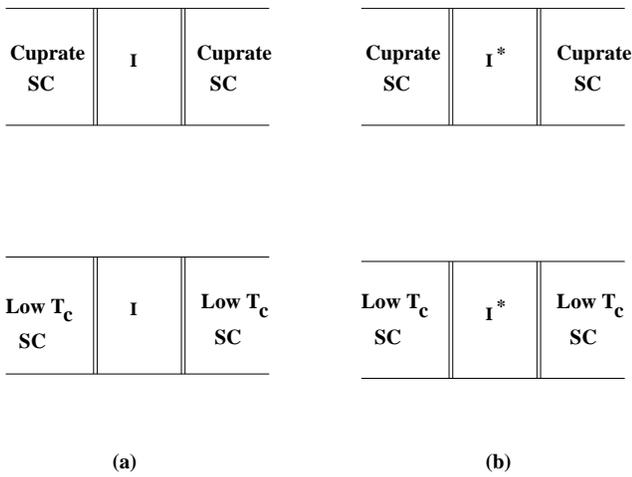}}
\vspace{0.15in}
\caption{Various kinds of Josephson 
junctions. In both cases in (a), the insulator $I$ is not fractionalized. Then, the Josephson
coupling occurs through Cooper pair tunneling and the ac Josephson frequency
is $2eV/\hbar$. In both cases in (b), the insulator $I^*$ is fractionalized.  
It is now possible for chargons to tunnel coherently between the two 
superconductors (independent of whether they are cuprate or low-$T_c$
superconductors), and the ac Josephson frequency will be $eV/\hbar$.}
\vspace{0.05in}
\label{ivsist}
\end{figure}

It is important to note that our result does {\em not} depend on 
the ``pairing" symmetry of the superconductor. In particular, the two 
superconductors could be conventional low-$T_c$ $s$-wave materials
(see Fig. \ref{ivsist}). 
This perhaps surprising fact further emphasizes our point that 
Cooper pair condensation and chargon condensation lead to the same 
superconducting phase. The charge has no integrity as a good quantum number
inside the superconducting state. Thus it is possible to halve 
the ac Josephson frequency to $\frac{eV}{\hbar}$ by making an insulating
barrier in which the chargons can freely propagate.

These general points are further illustrated by considering an
insulating barrier made of a conventional material in which the electron is
{\em not} fractionalized. Now chargons can no longer propagate
coherently from one superconductor to the other. However, the Cooper pair 
tunneling can proceed as before. Formally, this may be seen in the 
$Z_2$ formulation by noticing 
that the non-fractionalized phases are obtained when $h >> K$. 
In this case, the operator $\prod_C \sigma^z_{rr'}$ will fluctuate
very rapidly with average value zero. The amplitude for single 
chargon tunneling is thus zero, and only Cooper pair tunneling occurs. 
We therefore then obtain the standard ac Josephson effect with
frequency $\frac{2eV}{\hbar}$. Once again, this result too 
holds independently of the pairing symmetry of the superconductor. 
In particular, this is true despite our description of the cuprate superconductors
as a charge $e$ condensate.

\begin{figure}
\epsfxsize=3.3in
\centerline{\epsffile{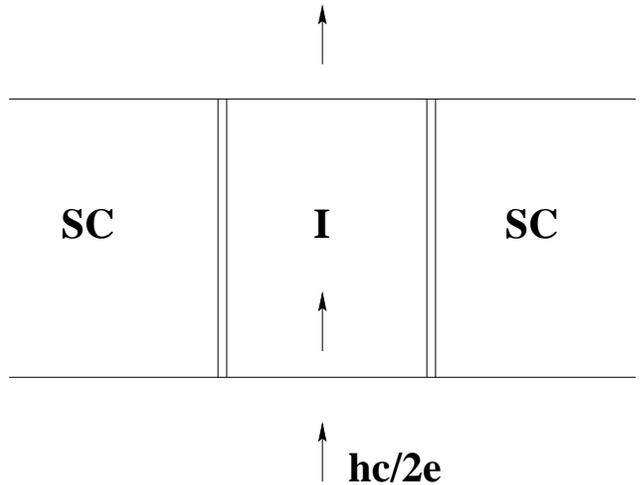}}
\vspace{0.15in}
\caption{The Josephson coupling in a conventional junction may be 
understood as due to slippage of $hc/2e$ vortices through the insulator.}
\vspace{0.05in}
\label{sisvtx}
\end{figure}

Further insight may be obtained by noting that the conventional Josephson effect may 
be thought of as due to a phase-slip process in which $\frac{hc}{2e}$
vortices pass through the insulator as shown in Fig. \ref{sisvtx}. 
If the insulator is fractionalized, 
then it allows free propagation of $\frac{hc}{e}$ vortices\cite{NLII,z2long}. 
But the $\frac{hc}{2e}$
vortices are gapped, and their propagation is suppressed. Indeed, the vison
is precisely the remnant of the gapped $\frac{hc}{2e}$ vortex in the fractionalized
insulator\cite{z2long}. Motion of an
$\frac{hc}{e}$ vortex across the junction corresponds to a chargon phase-slip by $2\pi$,
or equivalently a Cooper pair phase slip by $4\pi$. This then leads to 
ac Josephson oscillations at frequency $\frac{eV}{\hbar}$.
If the lateral width $w$ 
of the junction is finite, then the visons will slip through the interface at a
rate that is exponentially small in $w$. This will then restore the 
conventional Josephson coupling at long time scales. Thus, the result in Eqn. \ref{Jos}
assumes the limit of large $w$ as mentioned earlier.

Another consequence of the possibility of
coherent chargon tunneling through the insulator is that if a dc SQUID
is made with fractionalized insulators for the barriers, the current
will be a periodic function of the flux enclosed with period $hc/e$ rather than the 
conventional $hc/2e$.

\begin{figure}
\epsfxsize=3.3in
\centerline{\epsffile{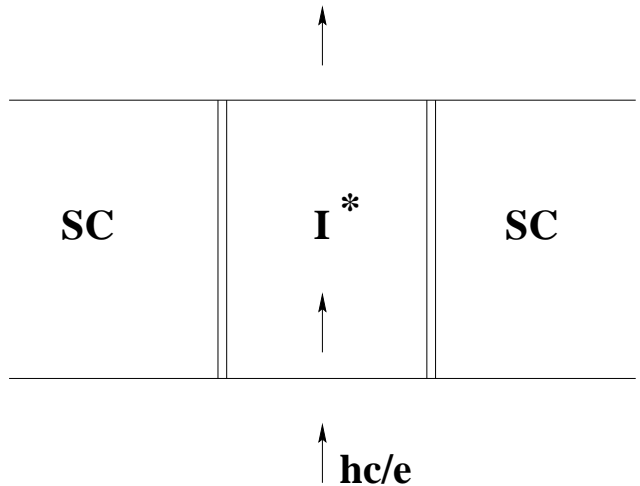}}
\vspace{0.15in}
\caption{The Josephson coupling with a fractionalized insulating barrier may be 
understood as due to slippage of $hc/e$ vortices through the fractionalized insulator. 
The $hc/2e$ vortices are not free to slip through the fractionalized insulator.}
\vspace{0.05in}
\label{sistvtx}
\end{figure}

Some words of caution are necessary in performing experiments to 
look for the anamolous ac Josephson effect in the undoped cuprate. 
First, if the cuprates are fractionalized, it seems most likely that the 
fractions of the electron are confined along the $c$-axis. In particular, 
the chargons cannot tunnel coherently between successive Copper-Oxygen layers. 
This implies that the anamolous ac Josephson effect will be seen only in 
the geometry in which the $c$-axis of the cuprate insulator is 
parallel to the interface.  
Second, thus far we have assumed a perfect interface between the superconductors 
and the insulator. If the interface is weak, so that the insulator is only 
weakly connected to one or both superconductors, the anamolous Josephson
effect will not be seen. This is because, as argued in Ref. \cite{topth}, 
along any line of weak contact, 
there will be a ($T=0$) phase transition at which 
coherent chargon (or spinon) tunneling across the line will
be blocked. Equivalently, along such a ``weak line", visons can slip
through unhindered - this will restore the standard Josephson effect. 
It is therefore necessary to have interfaces that are good enough that 
vison slippage through the interface is prevented as it is elsewhere 
in the junction.

\section{Coulomb blockade (``Dual'' Little-Parks)}

The hallmark of a superconductor is fluxoid quantization,
which follows directly from the condensation of a charged
boson.  In a neutral superfluid, it is simply the vortex 
which is quantized.
On the other hand, insulating states are characterized
by a quantization of the electric charge.  A direct way to measure
this quantization is by exploiting the Coulomb blockade
effect.  In a typical geometry a small metallic ``grain"
is electrically isolated from two metallic leads
by the presence of two insulating tunnel barriers.
Upon tuning the voltage on a gate electrode, $V_g$, conveniently
located to capacitively couple into the metallic grain
with capaciatance $C$,
it is possible to ``charge up" the grain one electron at a time.
This single electron charging can be detected by measuring the
elecrical conductance through the grain as a function of the
gate voltage.  One finds a periodic sequence of conductance
peaks with spacing $\delta V_g = e/C$ - each peak occuring
when there is a degeneracy between having $n$ and $n+1$ electrons
on the grain.  This Coulomb blockade
experiment can be correctly thought of as the
``dual" of the classic Little-Parks experiment - under the interchange
of flux with charge.

Exploiting the Coulomb blockade to detect possible electron fractionalization
in the underdoped cuprates is problematic since the
chargon fragment carries the full electron charge.  But as we now discuss,
it should nevertheless be possible if the small metallic grain is
replaced by a small superconducting grain.  
Coulomb blockade experiments involving a small superconducting
grain connected to metallic leads via two insulating tunnel
barriers have revealed\cite{cb} an astonishing ``even-odd" effect.
Due to the singlet pairing of electrons on the superconducting
grain, adding an extra electron to a grain with an
{\it odd} number of electrons is slightly less
costly (the gap energy) than  when the grain has an even number
of electrons.
This leads to an observable even-odd effect in the spacing between successive
conductance peaks, with the period set by the
distance between two peaks: $\delta V_g = 2e/C$ - 
a charge $2e$ periodicity corresponding to Cooper pairs
of electron.

\begin{figure}
\epsfxsize=3.3in
\centerline{\epsffile{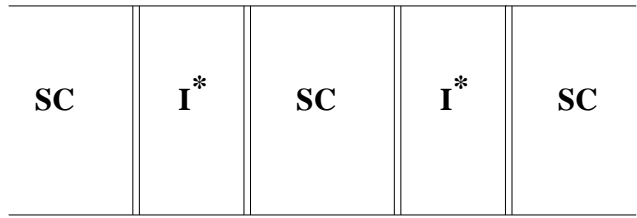}}
\vspace{0.15in}
\caption{Schematic of the Coulomb blockade experiment proposed to detect the chargon
in the fractionalized insulator. The superconducting island in the middle is separated
by fractionalized insulating tunnel barriers from the two superconducting leads. All
junctions are assumed to be perfect.}
\vspace{0.05in}
\label{coulbl}
\end{figure}

To detect the chargon,
we propose redoing this
Coulomb blockade experiment, 
making the insulating barriers from undoped cuprate
material.  Specifically, imagine a small superconducting grain
which is electrically isolated from two superconducting
leads with tunnel barriers made from undoped cuprate - as depicted schematically in Fig \ref{coulbl}.
The ideal experiment involves using conventional $s$-wave low $T_c$ superconductors\cite{note}. 
We presume that with opaque
barriers and a small grain, there is no 
Josephon coupling between the two electrodes.
Nevertheless,
each of the two barriers can be viewed as a Josephson junction
connecting the grain to the external leads, with a Josephson coupling energy
of the general form given in Eqn. \ref{Jos}.  If the undoped cuprates are
deep within a fractionalized phase, the single chargon hopping term
proportional to $t_1$ will dominate the pair tunnelling term.
In that case, as one tunes a gate potential which is capacitively
coupled to the grain, it should be a chargon
which is discretely hopping onto the grain - not
an electron or a Cooper pair.  Since the superconducting grain
is a chargon condensate (even
for a low $T_c$ material), these chargons can be 
readily absorbed by the condensate.  This implies
that the {\it conductance peaks should be charge-$e$ periodic}
- $\delta V_g = e/C$ - with no even-odd effect present. As in the discussion of the 
Josephson effect, here too it is necessary to have very good interfaces so that single
chargons can move freely through.

The halving of the $2e$ charge periodicity
when the insulating barriers are made from a fractionalized insulator,
is indicative of a ``vortex pairing"\cite{NLII,z2long}.  Specifically,
a fractionalized insulator descends from a conventional
superconductor when two $hc/2e$ vortices pair and condense.
The resulting $hc/e$ vortex-pair condensate
leads directly to halving
of the charge $2e$ periodicity
on the superconducting grain. This is the dual analog to
the halving of the electron flux quantization in
the original Little-Parks experiment.

\section{Stable $hc/e$ vortices}
Several years ago, Sachdev, Nagaosa and Lee\cite{SNL} pointed out the 
possible stability of $hc/e$ vortices in the superconducting
state close to the transition to a spin-charge separated normal phase. 
A single $hc/2e$ vortex is still a stable object, but a pair of them 
have higher energy than an $hc/e$ vortex. Here, we review the 
physics behind this observation, and use the currently available 
data to estimate the region of stability of the $hc/e$ vortex. 

The energy of any vortex in the superconducting state has two contributions.
First there is the energy of the superflow. This is determined by the 
superfluid stiffness $J$ - the coefficient of the 
$\frac{\left(\nabla \varphi \right)^2}{2}$ term in the Landau-Ginzburg
free energy for the phase of the Cooper pair. For a vortex of strength
$\frac{nhc}{2e}$, the superflow energy is given by
\be
E_{sf} = \pi n^2 J \ln\left(\frac{\lambda}{\xi} \right) ,
\ee
with $\lambda, \xi$ being the penetration depth
and the coherence lengths, respectively.   
The second contribution is the energy in the core of the vortex. 
If the underdoped cuprates emerge from a normal state
that is fractionalized, the $hc/2e$ vortex is made possible by 
the presence of the vison. Thus, the core energy of an $hc/2e$ 
vortex includes the energy cost of a vison. Now let us assume that 
the very underdoped insulator is fractionalized. Then the vison 
gap is non-zero in the insulator, and is expected to be smooth
across the superconductor-insulator transition. Consequently, the 
core energy of the $hc/2e$ vortex inside the superconducting 
state in the very underdoped regime 
is roughly the same as the vison gap, and is non-zero on approaching the 
quantum transition to the insulator. On the other hand, the $hc/e$
vortex does not have a vison in it's core and it's core energy 
vanishes on approaching the superconductor-insulator transition
assuming it is second order. 
(This is also consistent with the idea that the fractionalized
insulator may be viewed as a condensate of $hc/e$ vortices). 
In our earlier work\cite{z2short}, we have suggested that the vison gap is 
roughly of the order $k_B T^*$ where $T^*$ is the temperature associated
with the pseudogap crossover. We may therefore estimate
\be
E^{core}_{\frac{hc}{2e}}  \simeq k_B T^*.
\ee

It is clear from the above that the difference between 
the energy of a single $hc/e$ vortex and
that of two well-separated $hc/2e$ vortices
is
\be
E^{core}_{\frac{hc}{e}} + 2\pi J \ln \left(\frac{\lambda}{\xi}\right) - 2 k_B T^*. 
\ee
If the $T=0$ transition from the superconductor to the fractionalized insulator is second order, then 
the core energy of the $hc/e$ vortex must vanish on approaching the transition. Further, we expect 
that this core energy will essentially be set by $J$ - thus it is numerically smaller than the 
superflow energy by a factor of order $\ln\left(\frac{\lambda}{\xi}\right) \approx 5$ (see below).  
For a rough estimate we drop it completely.
Thus, for the $hc/e$ vortex to be cheaper, we need
\be
\pi J \ln \left(\frac{\lambda}{\xi}\right) \approx k_B T^*.
\ee
Clearly, this will always happen close enough to the 
transition. Empirically, the zero temperature stiffness
is proportional to $k_B T_c$ in the underdoped regime. 
In $YBCO$, we have\cite{lambdatotc} $J(T = 0) \approx 1.4k_BT_c$. Further, we have
$\lambda \approx 1600 \AA$, $\xi \approx 10 \AA$. 
We thus have the rough condition
\be
7\pi T_c \approx T^* ,
\ee
on the maximum $T_c$ for $hc/e$ vortex stability.  

The estimate above addresses the issue of the stability of the $hc/e$
vortex at zero temperature. On moving up in temperature, the 
stiffness $J$ decreases thereby decreasing the 
superflow contribution while there should be no significant change in the 
core energies. Thus, the $hc/e$ vortex would gain in stability. 

Considerable caution is required in trying to observe these stable 
$hc/e$ vortices in experiments. The force between two $hc/2e$ vortices is
always repulsive at large separation (much bigger than the core size) where it is dominated
by the superflow. Thus it is necassary for two well-separated $hc/2e$ vortices to
overcome the superflow energy barrier and get close enough before the gain 
in core energy of the $hc/e$ vortex can provide for the attraction to bind them together. 
In practice, depending on the dynamics and the history of the sample, it may be possible
for $hc/2e$ vortices to be observable in some highly metastable state 
even in a regime in which a single $hc/e$ vortex has lower energy than a pair of 
$hc/2e$ ones. 
 
\section{Discussion}
In this paper, we have proposed a number of tests of the 
idea that the electron is fractionalized in the underdoped and undoped insulating
cuprates. 
These experiments are sensitive to the presence of the chargon in the excitation spectrum 
of the insulator. As such, they allow for a direct experimental 
probe of the question of 
whether the undoped cuprate is fractionalized. In our earlier work\cite{toexp,topth}, 
we proposed an
experiment to detect the vison in the underdoped cuprates. This vison detection
experiment is however not very suitable to addressing the issue of 
fractionalization in the undoped cuprate. The experiments
proposed in this paper are thus complementary to this vison detection experiment.
Taken together, we believe that a positive result in any of these experiments 
would be compelling evidence for fractionalization of the electron in the 
cuprates. 

We conclude by reemphasizing one intriguing aspect of 
the results in this paper. 
We have argued that in both the ac Josephson effect and in the Coulomb blockade
tunneling experiment, the outcome depends 
sensitively on whether the insulating tunnel 
barrier is fractionalized or not. Surprisingly, the precise nature (``pairing"
symmetry) of the 
superconducting state is unimportant. In particular, both experiments 
should be possible with conventional low-$T_c$ superconductors.
This implies that it is not always convenient
to view even a low $T_c$ superconductor as a Cooper pair condensate.
Rather, for the experiments described here, 
the superconductor is best thought of as a {\it chargon} condensate. 
This ambiguity in the ``charge of the condensate"
is due to the fact that the charge is not a  
good quantum
number in a superconductor, so that Cooper pairs
(or chargons) do not  {\em really}
exist inside any superconductor!

We are particularly grateful to Kathryn Moler for her insightful questions which led to 
several of the results of this paper. 
We also thank L. Balents, J. Berlinsky, D. Drew, C. Kallin, P.A. Lee, T. Leggett, 
and S. Sachdev for useful discussions. This research was supported by the NSF 
under Grants DMR-97-04005,
DMR95-28578
and PHY99-07949 .

\end{multicols}

\begin{references}
\bibitem{note} If the superconductor was $d$-wave, then it would have gapless
nodal quasiparticle excitations. Tunneling of these could give rise to charge 
$e$ periodicity even with a conventional (unfractionalized) insulating tunnel barrier. 
Thus, the absence of an even-odd effect will not establish the presence of 
fractionalization in the insulator.
\end{references}
\end{document}